\documentclass[preprint,12pt]{elsarticle}
\usepackage{graphicx,amssymb,amsmath,epsf,color,hyperref}
\usepackage{epstopdf}
\usepackage{enumitem}   
\usepackage[dvipsnames]{xcolor}
\usepackage{amsmath}
\usepackage{amssymb}
\usepackage{hyperref}
\usepackage[para]{threeparttable}

\begin{document}

\begin{frontmatter}

\title{Different Environmental Conditions in Genetic Algorithm}
\author{Daekyung Lee}
\author{Beom Jun Kim\corref{cor1}}
\ead{beomjun@skku.edu}
\cortext[cor1]{Corresponding author}
\address{Department of Physics, Sungkyunkwan University, Suwon 16419, Republic of Korea}

\begin{abstract}

We propose an extended genetic algorithm (GA) with different local
environmental conditions.  Genetic entities, or configurations, are put on
nodes in a ring structure, and location-dependent environmental conditions are
applied for each entity.  Our GA is motivated by the geographic aspect of natural
evolution: Geographic isolation reduces the diversity in a local group, but at
the same time, can enhance intergroup diversity. Mating of genetic entities
across different environments can make it possible to search for broad area of
the fitness landscape.  We validate our extended GA for finding the ground
state of three-dimensional spin-glass system and find that the use of different
environmental conditions makes it possible to find the lower-energy spin
configurations at relatively shorter computation time. Our extension of GA
belongs to a meta-optimization method and thus  can be applied for a broad
research area in which finding of the optimal state in a shorter computation
time is the key problem.
\end{abstract}
\end{frontmatter}

\section{Introduction}
%\label{Sec:intro}
%{\it Introduction$-$}
Genetic algorithm (GA) has been proposed by John H. Holland in 1975~\cite{holland}, and has become a 
well-adopted technique in various research areas to solve computationally difficult problems~\cite{goldberg}.
In the process of Darwinian evolution, various mechanisms of genetic mutations
produce offsprings of different fitness, and the fitter tend to survive and produce more offsprings later on.   
Likewise, GA in various optimization problems first produces a group of candidate solutions, and only a part
of them are then selected to become parent solutions to next generation. 

In the most basic scheme of GA, each of possible solutions to the target problem is encoded as a genotypic entity or 
a configuration, and a set of initial genotype populations are generated randomly and ranked in terms of the {\em fitness}, a 
specific variable which denotes how well the individual suits the conditions of the target problem. 
Each individual entity has a chance to mate with another entity and to produce offsprings, whose genetic 
traits are inherited from parents. The entity with a high fitness is set to be advantageous 
and the system as a whole tends to approach the desired optimum with the highest fitness value. 
It is a widespread observation that GA can locate efficiently 
an approximate solution of a given complex problem in a relatively short time~\cite{perfor1,perfor2,perfor3}.

However, despite of its usefulness, basic GA has a few drawbacks which have to
be resolved. For example, since genotypes of offspring entities are generated
by the combination of genotypes of their parents, a simple GA system 
possesses a risk of losing genetic diversity and converging into only a few genotypes
near a local maximum in the fitness landscape. 
Most of GA's use the mutation 
process, which adds a random noise to genotypes of each offsprings, to maintain a certain level of the genetic 
diversity~\cite{Diversity}. In the same spirit, many revised or extended mechanisms have been suggested to improve various aspect of 
GA's: Wiriyasermkul {\it et al.} have proposed a meiosis GA, which imposes a duplication process to create two 
types of offsprings~\cite{meiosis}. S{\' a}nchez-Velazco {\it et al.} assigned a gender attribute to the genetic entity
and introduced the effect of asymmetric mating and reproduction process~\cite{gendered}. 

In the present paper, we propose an extension of GA which utilizes a geographic
aspect of natural evolution~\cite{vari1,vari2,vari3}.  Such geographic effect
can be easily seen in the famous example of Darwin's finches: Different species
of finches evolved in different local environments. We are motivated by such
well-known observation in nature, and propose that migrations and mating
across different local environments can broaden the search space of the fitness
landscape. Location-specific environment allows largely different genetic
entities to survive in different local conditions and mating between genetic
entities at separate locations can provide a larger diversity in a gene pool.

Our GA needs to be compared with the island model~\cite{island_origin}, in
which genetic entities are distributed across islands. Although mating is
allowed only within an island, islands are allowed to exchange genetic entities
with each other, and thus successful offsprings can spread out.  The island
model as well as our GA in the present paper can be phrased as a
meta-optimization algorithm since one can apply suggested methodology on top of
the conventional optimization algorithm. In the island model, the frequency and
the magnitude of the exchange across islands are tunable parameters for the
meta-optimization, and our GA uses the difference in environmental conditions
for that purpose.  It has been reported that properly adjusted island model
often outperforms the basic GA~\cite{island}. Likewise, we hope to optimize our
modified GA with different environmental conditions to achieve better performance.  
We emphasize that the present work does not aim to get the best optimum for
a given problem. Instead, we are trying to show that addition of different
environmental conditions for a given optimization problem can be helpful
in a practical sense to find a better optimum in a shorter computation time.

In our approach, genetic entities are organized into a specific network
structure so that each of them can have a regional attribute. By applying
heterogeneous environmental factors to a suitably defined fitness function, we
are able to construct a genetic system which consists of individuals with
different location-specific objectives. If each genetic entity is allowed to
mate with a local or a remote partner, we are able to observe how
environmental difference between parents affects the survivals of offsprings. It
is to be noted that we use the geographic isolation in a positive way: 
Isolation makes each local group lose
genetic diversity within the group, but at the same time, it can help each
local group to have a genotype largely different from other local groups due to
different environmental conditions. In other words, our GA with heterogeneous
environmental factors reduces intragroup diversity but enhances intergroup
diversity, and the latter is the main source of genetic diversity in our
framework.

\section{Method}

In order to investigate the concrete effect of our GA, we use a specific
physics problem of finding the ground state of the Edwards-Anderson spin-glass
model~\cite{ed-an} with an external field. we again emphasize that
we are not seeking the lowest possible energy state of the model, but we hope
to show that the use of the different environmental condition can effectively
enhance the performance of GA in finding lower-energy state. 
The EA model has been widely used as an archetype 
to investigate complex collaborative glass behavior. The Hamiltonian of the EA model reads  
\begin{equation}\label{eq:Hamil1}
\mathcal{H} = -\sum_{\langle i j\rangle}J_{ij}s_{i}s_{j} - h\sum_{i}s_{i},
\end{equation}
where the sum $\sum_{\langle i j \rangle}$ runs only for nearest-neighbor pairs with
the periodic boundary condition, $s_i (=\pm 1)$ is the Ising spin variable at the $i$th site, $J_{ij}$ is the coupling 
strength of the atomic bond between spins at the sites $i$ and $j$,  and $h$ is the spatially uniform external field.
Note that the local interaction structure makes it difficult to obtain the ground state of the system: It has been 
proven that finding the ground state of the EA model in three dimensions (3D) is NP-complete problem~\cite{NP}. 
Of course, we can instead use a numerical approach to find the ground state, and several 
techniques, such as 
simulated annealing~\cite{simul}, population annealing~\cite{popu}, extremal optimization~\cite{EO}, 
genetic algorithm~\cite{E1D}, and branch-and-cut algorithm~\cite{exact} have been suggested and applied in 
previous researches. In particular, the branch-and-cut algorithm is known to find the exact ground state 
in shorter computation time\cite{exact}. Although it is challenging work to make better algorithm than existing ones 
designed for the specific problem, we rather focus on more general aspect of optimization problem. In the present work we aim to propose how one can enhance the performance
of a genetic algorithm through the environment-driven genetic diversity with the 3D EA model as
an exemplar system.

\begin{figure*}
\includegraphics[width=\textwidth]{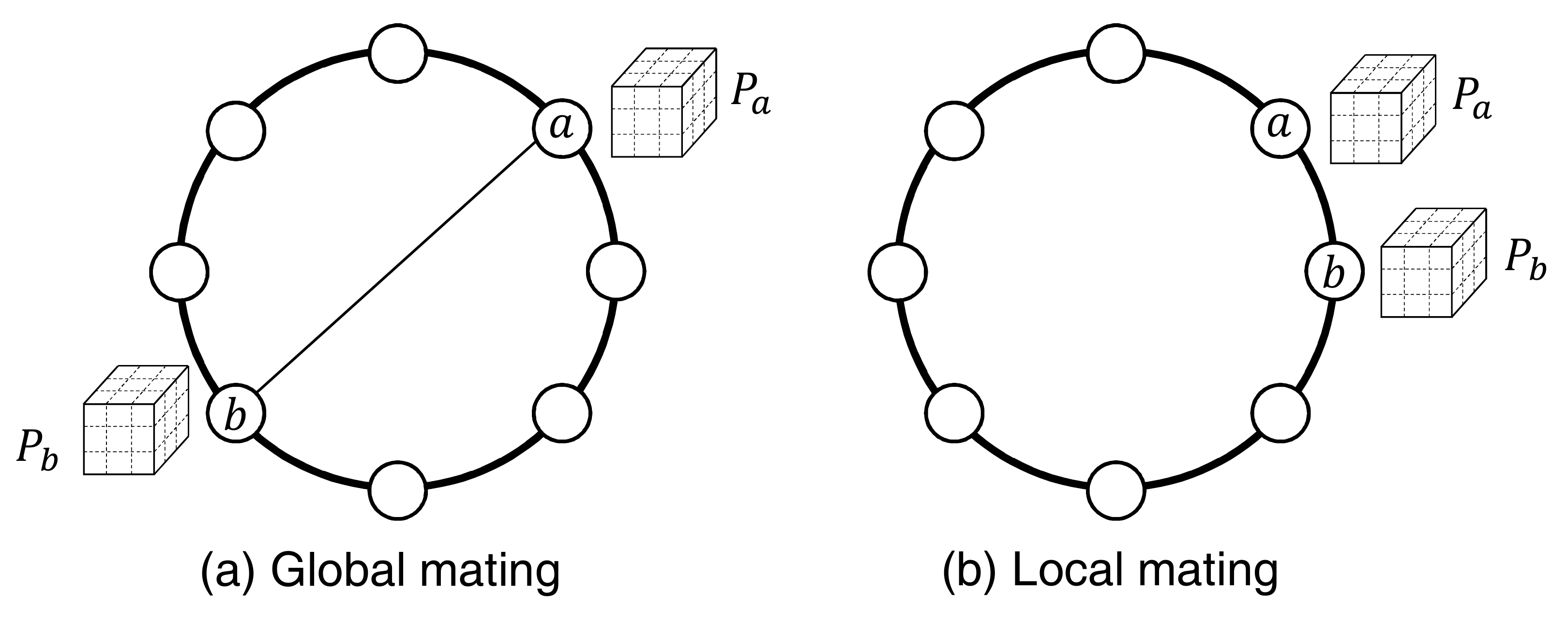}
\caption{Schematic illustration of our genetic algorithm. A ring network of $M$
nodes is constructed, and a spin system is put on each node. In more general
context, a node can represent a genetic entity that evolves in time. (a) Global
mating: At the global mating probability $p$ we pick a node at $b (\neq a)$ completely at random
among all other nodes in the ring network. (b) Local mating: with the probability
$1-p$, one (at $b$) of $a$'s two nearest neighbors is randomly chosen.
The two spin systems $P_a$ and $P_b$ at $a$ and $b$, respectively, mate and
reproduce offsprings (see text). }
\label{fig:mating}
\end{figure*}

The basic scheme of our genetic algorithm starts from a ring network which consists of  
$M$ nodes as shown in Fig.~\ref{fig:mating}. Each node represents a whole three-dimensional spin configuration of the linear size $L$ (and thus the total number $N$ of spins satisfies $N = L^3$), which is connected to 
two (left and right) neighbor nodes as in Fig.~\ref{fig:mating}. The node index $a$ increases in the counterclockwise direction from 1 to $M$. Since nodes are arranged in the form of a ring, the distance between two nodes $a$ and $b$ is  written as
$\min(M-|a-b|,|a-b|)$, to impose the periodicity along the circular ring
structure. To emphasize that a node in the ring represents a whole 3D
spin-glass system, we call each node as {\it a system node} from now on. 

The Hamiltonian of the system node $a$ is defined according to Eq.~(\ref{eq:Hamil1}) as
\begin{equation}\label{eq:Hamil2}
\mathcal{H}^{a}(t) = -\sum_{\langle i j\rangle}J_{ij}s_{i}^{a}s_{j}^{a} - h^{a}(t)\sum_{i}s_{i}^{a},
\end{equation}
where $s_{i}^{a}$ is the $i$-th spin of the system node $a$ and satisfies the
periodic boundary condition ($s^a_{i+L \hat x} = s^a_{i+L \hat y} = s^a_{i+L
\hat z} = s^a_i$ with $\hat x$, $\hat y$, and $\hat z$ being unit vectors in
the 3D Cartesian coordinate system), and the coupling strength $J_{ij}$ is the
quenched random variable generated from the Gaussian distribution with zero
mean and unit variance.  It is important to note that  $\{ J_{ij} \}$ is
identical across system nodes, and thus all system nodes are assigned exactly
the same optimization problem in the absence of the external field.  
The external field $h^{a}(t)$ explicitly depends on the node
index $a$ and the time $t$, which is the key ingredient in our GA to mimic the different
environmental condition in the evolution process.  
The time variable $t$ can be considered as an index for generation,
to be explained more below.  

In order to implement a geographically heterogeneous and time-varying environmental condition, 
the external field $h^{a}(t)$ with the amplitude $h_0$ is written as 
\begin{equation}\label{eq:field}
h^{a}(t) = h_{0}\sin \left[ 2\pi \left( \frac{a}{M} - \frac{t}{T} \right) \right],
\end{equation}
which is periodic both in spatial and temporal dimensions.  
The spatial periodicity $h^a(t)$ can
be interpreted as, e.g., the change of environmental condition with the latitude
on the earth, whereas the temporal periodicity of $h^{a}(t)$ can be interpreted
as some long-term climate change like the periodic occurrences of ice
ages on the earth. Since $h^a(t)$ oscillates in time around zero, each system
node faces the same optimization problem in an average sense. Note that both 
parameters affect the fundamental dynamics of a genetic system and we use them 
as parameters in the meta-optimization scheme.

Our genetic algorithm in the present work can be roughly divided into five
stages: (i) initialization, (ii) mating, (iii) crossover, (iv) mutation, and
(v) competition. We describe each stage one by one in the followings.
\begin{enumerate}[label=(\roman*)]
\item {\em Initialization}: All spins at each system node in the ring network are
randomly set to $1$ or $-1$ at equal probability. 
\item {\em Mating}: We randomly pick a parent $P_a$ at the system node $a$.
The other parent $P_b$ at the system node $b$ to mate with $P_a$ is chosen
in either of the two ways: the global mating with the probability $p$ and the local mating 
with probability $1-p$ (see Fig.~\ref{fig:mating}).
For the former, $b$ is chosen at random among all other system nodes,
while for the latter mating, $b$ is chosen between the two (the left and the right) 
neighbor nodes of $a$. Throughout the present work, we use the global mating probability
$p = 1/2$.
\item {\em  Crossover}: Spins of two parents $P_a$ and $P_b$ 
are combined to create the first ($O_1$) and the second ($O_2$) offsprings
 whose spin configurations 
are written as $\{ s_i^{O_1} \}$ and $\{ s_i^{O_2} \}$, respectively, 
with $i=1, 2, \cdots, N$.
We assume that the crossover process is uniform across all spins,
and thus there are only two possibilities for each spin at the $i$th site:
$(s_{i}^{O_1} = s_i^{P_a}, s_{i}^{O_2} = s_{i}^{P_b})$ or $(s_{i}^{O_1} =
s_i^{P_b}, s_{i}^{O_2} = s_{i}^{P_a})$ at equal probability of 1/2. 
Note that this crossover process we adopt here is the most basic one, 
and that there are other more advanced schemes~\cite{crossover}. 
\item {\em Mutation}: Each spin in the offspring configurations is reversed ($s_i^{O_{1}} \rightarrow -s_i^{O_{1}}$, $s_i^{O_{2}} \rightarrow -s_i^{O_{2}}$) at the mutation probability $\mu = 0.01$. Although mutation process
has often been utilized to provide a genetic diversity in most previous studies, 
it is not the only source of genetic diversity in our framework. 
\item {\em Competition}:  
Once the spin configurations of two offsprings are constructed in (iv),
we then decide where these offsprings are to be born. There are two
possibilities: $O_1$ is at the system node $a$ and $O_2$ at $b$, and vice
versa. In order to maintain the continuity of genetic system, we make a reasonable 
assumption that the local environmental condition at $a$ will be more friendly for an 
offspring who has a spin configuration closer to the parent at $a$.
We use the Hamming distance $D(\alpha,\beta)
\equiv \sum_i [1 - \delta(s_i^\alpha, s_i^\beta) ]$ with the
Kronecker-$\delta$, to measure the similarity between two spin configurations
$\alpha$ and $\beta$.  We then compare $D(P_a,O_1) + D(P_b, O_2)$ and
$D(P_a,O_2) + D(P_b, O_1)$: If the former is less than the latter, this can be
interpreted as that $O_1$ and $O_2$ have spin configurations closer to those of
parents at $a$ and $b$, and thus we decide that $O_1$ is born at $a$ and $O_2$
at $b$. Otherwise, if  $D(P_a,O_1) + D(P_b, O_2) > D(P_a,O_2) + D(P_b, O_1)$,
the situation is reversed and thus $O_1$ and $O_2$ are born at $b$ and $a$,
respectively. At each location, we compute the energy values for the parent and
the new born offspring, and the one with the lower energy is selected to
survive there.  
\end{enumerate}
%In each stage,
%to assure the minimum effectiveness, we employ the basic framework 
%in~\onlinecite{E1D}.

\section{Results}
%\label{sec:results}
%{\it Results$-$}

In our simulations, the ring network (see Fig.~\ref{fig:mating}) has the size 
$M=100$, and each system node consists of the 3D EA spin-glass model of 
the linear size $L=5$ (and thus total number of spins $N = L^3 = 125$).
The two control parameters are the amplitude $h_0$ and the temporal period $T$
of the local magnetic field in Eq.~(\ref{eq:field}) in our meta-optimization scheme.

We initialize the spin configuration of each system node in (i) {\em
Initialization} step, and iterate the procedure (ii)-(v).  
We measure the time $t$ in such a way that steps (ii)-(v) are iterated once for
each system node on average in one unit of time. In other words, $M(=100)$ repetitions of steps
(ii)-(v) amount one unit of time. Simulations are stopped at $t = t_{\rm max} = 5000$,
which is long enough for most parameter values. We then repeat the above whole
procedure 5000 times with different random configuration of $\{J_{ij}\}$ and
averages are taken for these independent runs. 
In order to investigate the performance of our GA, we measure the minimum energy and 
average diversity at time $t$ defined by 

\begin{equation}
\label{eq:Eavg}
E_{\rm min}(t) \equiv -\left\langle \min_{a} \frac{1}{N}\sum_{\langle i j\rangle}J_{ij}s_{i}^{a}(t)s_{j}^{a}(t) \right\rangle , 
\end{equation}

\begin{equation}
\label{eq:Davg}
D_{\rm avg}(t) \equiv \left\langle \frac{1}{M(M-1)N}\sum_{b \neq a} 
\sum_i \Bigl\{ 1 - \delta\bigl[s_i^a(t), s_i^b(t)\bigr] \Bigr\}
\right\rangle ,
\end{equation}
where $\langle \cdots \rangle$ represents the ensemble average over 
iterations for 5000 different configurations of $\{ J_{ij} \}$.

\begin{figure}
\includegraphics[width=\textwidth]{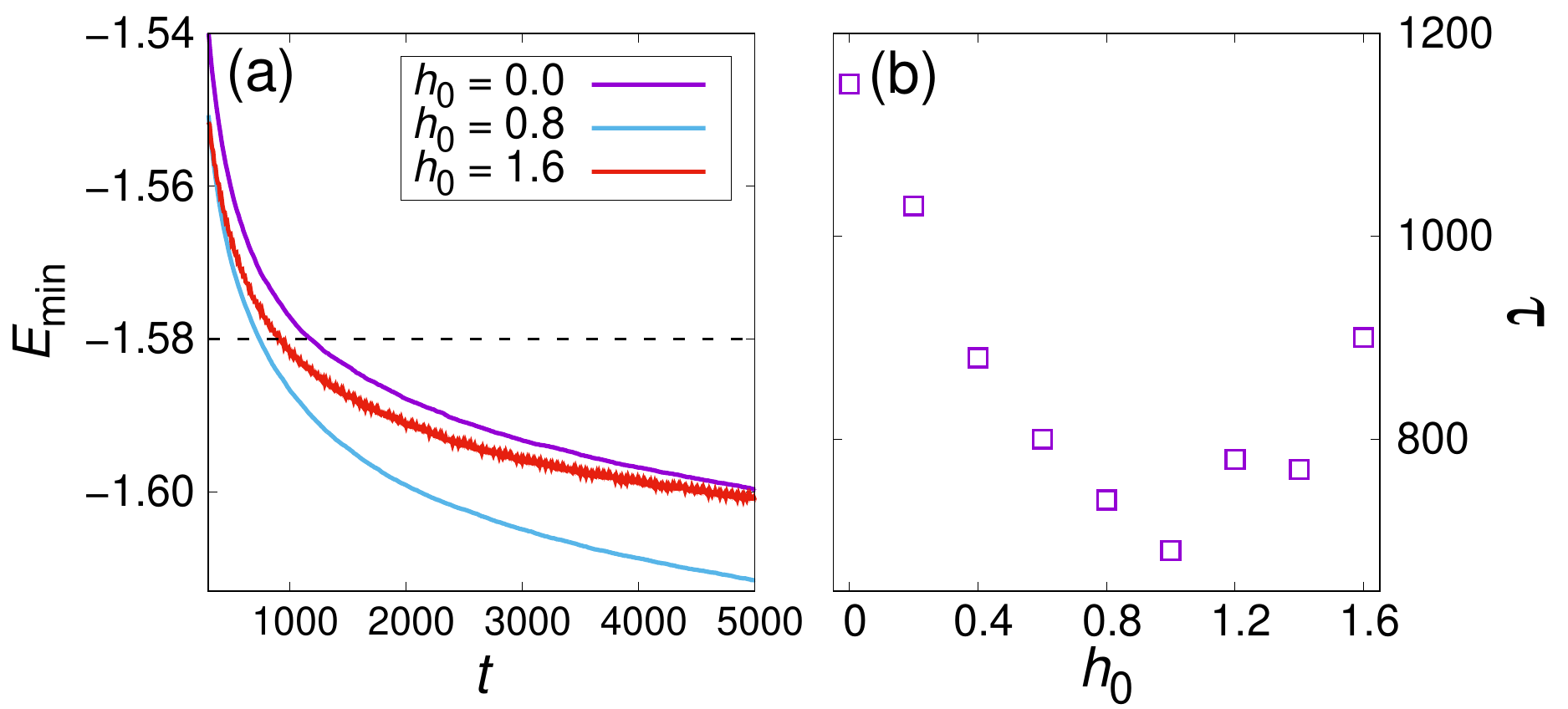}
\caption{(a) The minimum energy $E_{\rm min}(t)$ at temporal period $T=50$, as a function 
of $t$ at the field strength $h_0$ = 0.0, 0.8, and 1.6. (b) The decay time scale $\tau$ defined
by $E_{\rm min}(t = \tau) = -1.58$ [see the horizontal dashed line in (a)] exhibits minimum around
$h_0 = 1.0$, which indicates that our GA performs better at the intermediate field strength
and thus can find lower energy configuration in a relatively shorter computation time. }
\label{fig:time_evolving}
\end{figure}

In Fig.~\ref{fig:time_evolving}, we display $E_{\rm min}(t)$ versus time $t$
for $t \in [300, 5000]$ at $h_0 = 0.0, 0.8$, and 1.6, and the time period is
fixed at $T=50$.  In Fig.~\ref{fig:time_evolving}(a), we first notice that in
the absence of the external field ($h_0 = 0$), $E_{\rm min}(t)$ displays a
relatively slow decay.  At larger values of the field strength ($h_0 = 0.8$ and
$1.6$), on the other hand, the minimum energy decrease faster in early time
region.  It is very interesting to observe in Fig.~\ref{fig:time_evolving}(a)
that the decay of $E_{\rm min}(t)$ is faster at $h_0 = 0.8$ than at $h_0 = 0.0$
and $h_0 = 1.6$. This strongly suggests that our modified GA becomes more efficient
at the intermediate field strength in finding the lower-energy state. 

In order to check more carefully the nonmonotonic behavior observed in Fig.~\ref{fig:time_evolving}(a),
we measure the decay time scale $\tau$ at which $E_{\rm min} = -1.58$ is crossed, as represented by
the horizontal dashed line in Fig.~\ref{fig:time_evolving}(a).  The smaller $\tau$ is, the
faster our GA approaches a lower-energy state.
Figure~\ref{fig:time_evolving}(b) exhibits
the decay time $\tau$ as a function of the field strength $h_0$ at $T=50$, which
clearly shows the expected non-monotonic behavior. We thus conclude that the use of the proper
strength of the heterogeneous field reflecting the different environmental condition can 
make our GA more efficient.

\begin{figure}
\centering
\includegraphics[width=0.75\textwidth]{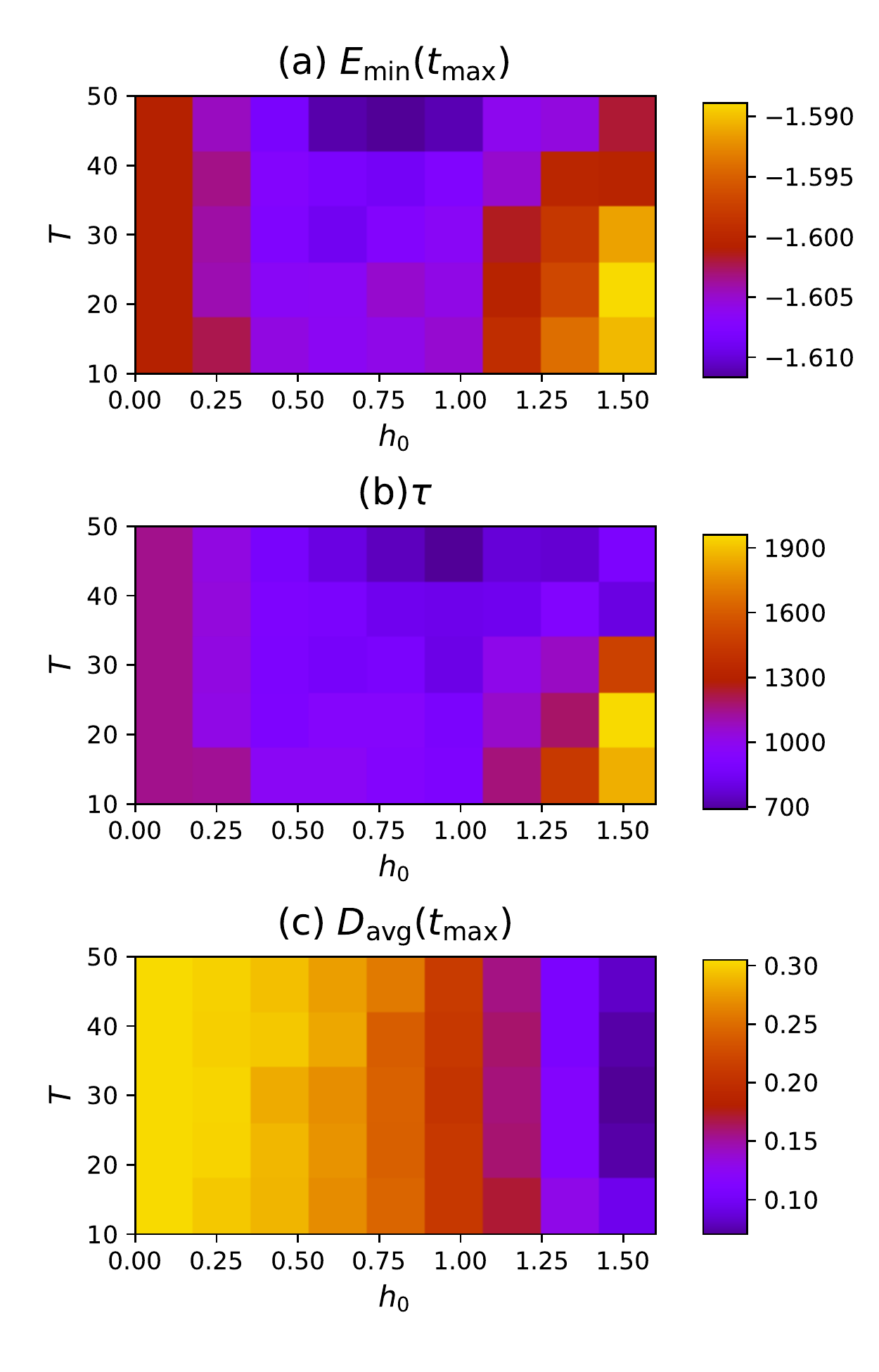}
\caption{(a) The minimum energy $E_{\rm min}(t_{\rm max})$, 
(b) the decay time $\tau$ defined by $E_{\rm min}(t = \tau) = -1.58$, and (c) the genetic 
diversity $D_{\rm avg}(t_{\rm max})$, with $t_{\rm max} = 5000$ in the plane of ($h_0$,$T$).
Note that the use of the field strength e.g., $h_0 \approx 0.8$ and the temporal period 
$T \approx 50$ can achieve a lower energy [see (a)]  in a shorter computation time [see (b)]
with preserving an appropriate level of genetic diversity [see (c)].}
\label{fig:heatmap}
\end{figure}

Figure~\ref{fig:heatmap} displays the
performance of our GA at various values of $h_0$ and $T$. 
We first notice that $E_{\rm min}(t_{\rm max} = 5000)$ in Fig.~\ref{fig:heatmap}(a) exhibits
a minimum around $h_0 \approx 0.8$ and $T \approx 50$.
As can be seen in Fig.~\ref{fig:time_evolving}(a), the system 
may not have approached a steady state till $t=5000$. 
However, we limit the simulation time up to $t_{\rm max} =5000$, to focus 
on practical applicability of our method: We aim to find low
enough energy state for a given limitation of simulation time.
As was already seen in Fig.~\ref{fig:time_evolving}(a), the observation
implies that a certain level of inhomogeneity in the environmental condition
is beneficial in finding the lower-energy state.

The behavior of the decay time $\tau$ in Fig.~\ref{fig:heatmap}(b)
also shows a similar tendency: The minimum occurs at $h_0 \approx  1.0$ and 
$T \approx 50$. In addition, the genetic diversity $D_{\rm avg}(t_{\rm max})$ 
measured at $t = t_{\rm max}$ in Fig.~\ref{fig:heatmap}(c) provides a validity of above results. 
Although the system loses its diversity as $h_0$ becomes larger, we confirm 
that significant amount of the diversity remains in the intermediate region of $h_0$ in comparison to
the null model at $h_0 = 0$.
Overall, we confirm that the use of a proper strength of the inhomogeneous external field 
can efficiently locate the lower-energy state in a shorter computation time 
when combined with sufficiently long temporal period $T$, still preserving a certain level of diversity.

\begin{figure*}
\includegraphics[width=0.9\textwidth]{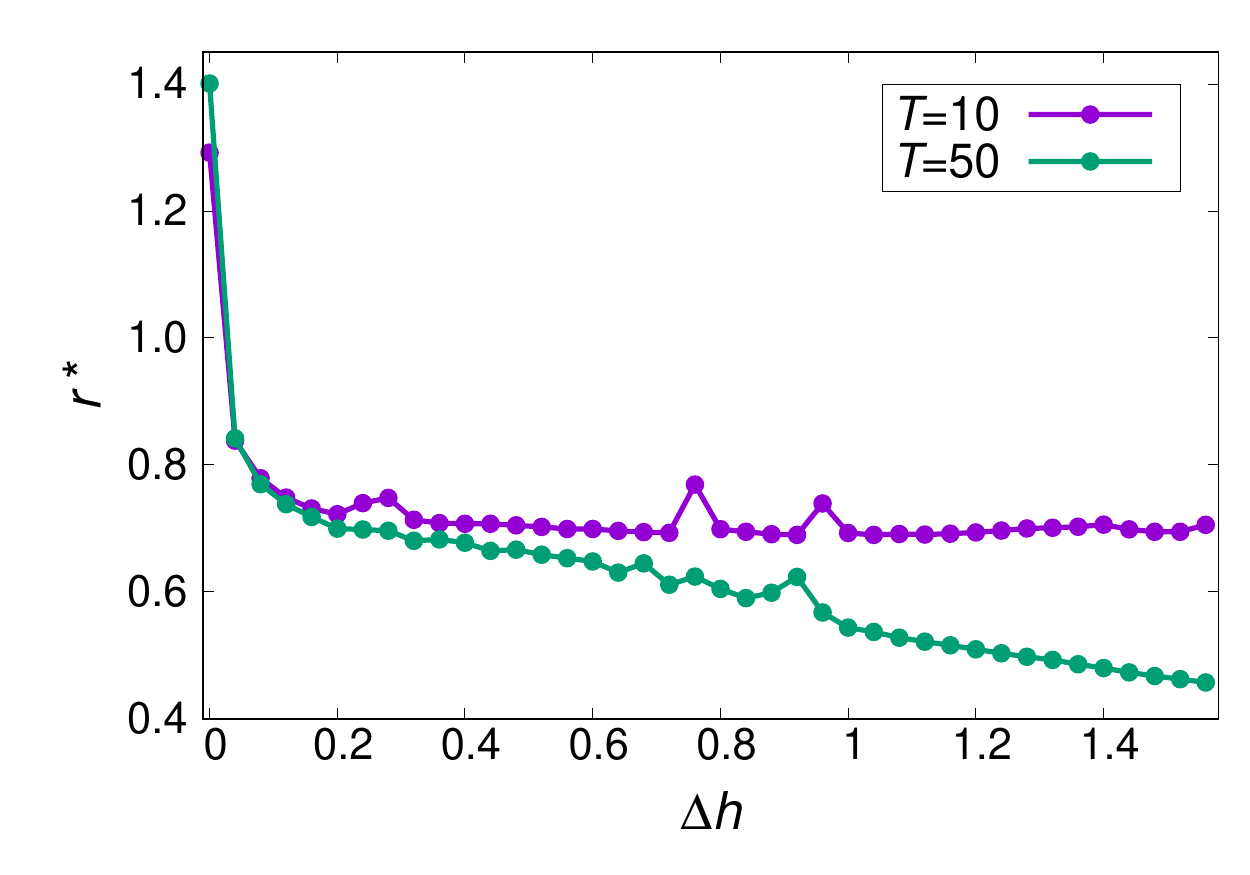}
\caption{Normalized acceptance ratio $r^*$ (with the bin size 0.04) versus the field
difference $\Delta h$ between two parent nodes for $h_{0}=0.8$ and $T=10, 50$.
The slow ($T=50$) and the fast ($T=10$) temporal changes of the external field
exhibit very different behaviors, which can be compared with Fig.~\ref{fig:heatmap}.
}
\label{fig:acc_ratio}
\end{figure*}

We next investigate the origin of the success of our proposed GA with 
focus put on the details of the mating stage.  
We measure the field difference $\Delta h
\equiv |h^{a}(t)-h^{b}(t)|$ for parent systems $P_a$ and $P_b$ at $a$ and $b$,
and compute the acceptance ratio $r$ of the offspring replacing the parent. 
For a better comparison, we use the normalized acceptance ratio 
$r^* \equiv r/r_{\rm avg}$ with $r_{\rm avg}$ being the time average of $r$
for $t \in [0, 5000]$. Figure~\ref{fig:acc_ratio} shows $r^*$ versus $\Delta h$ 
for  $T=10$ and 50 at $h_{0}=0.8$. We have chosen the two values $T=10$ and 50
to compare the difference between the fast and the slow temporal change (see Fig.~\ref{fig:heatmap}). 
The region of relatively small values of $\Delta h$ in Fig.~\ref{fig:acc_ratio} 
mostly reflects what happens in the local mating, while larger values of $\Delta h$
are from the global mating. 

In Fig.~\ref{fig:acc_ratio}, $r^*$ for $T=50$ tends to decrease monotonically with $\Delta h$
while $r^*$ for $T=10$ appears to be insensitive to the change of $\Delta h$ in a broad region. 
Such a difference, we believe, indicates that a sufficiently large value of $T$
is required for the better performance of our method. In other words,
only when the temporal change of the external field is sufficiently slow, 
our GA method becomes efficient to enjoy the benefit of the different environmental conditions
across the whole system, yielding the better performance observed in Fig.~\ref{fig:heatmap}. 
Consequently, we conclude that our GA exhibits an improved performance when the system 
has enough amount of spatial difference, which counteracts the innate tendency of 
homogenization in the conventional genetic algorithm.

%When a genetic system conducts a local mating, each node collaborates with its
%neighboring nodes so that efficiently finds the solutions which are adequate in
%the field condition at its vicinity. While, in the global mating process, each
%node is able to mate with foreign nodes which adapted in substantially
%different field condition. 

\section{Summary and Discussion}
%\label{sec:discussion}

In this paper, we have proposed a genetic algorithm with the heterogeneous
environmental condition, motivated by the effect of the geographic and temporal
variation of environmental condition in natural evolution. Different
environmental conditions can reduce intragroup diversity, but can enhance
intergroup diversity.

We have used the three-dimensional spin-glass system as a genetic entity, and thus the
lower-energy state of the system corresponds to the higher fitness in evolution.  
Our simulation results suggest that the existence
of a certain degree of inhomogeneous and time-varying external field can help
us to find a lower-energy state in a shorter computation time when the temporal
period of the field is sufficiently large.

We again emphasize that in the present work we do not intend to propose 
the most efficient algorithm in finding the ground state of the spin-glass problem in particular. 
Even though our result in previous section only deals with a relatively higher energy regime $\approx$ -1.61 
in comparison to the known ground state of target system $\approx$ -1.73\cite{E1D}, 
we use the spin-glass system only as an example to validate the general applicability of our method in a broad range 
optimization problems. In other words, we believe that our proposed genetic algorithm belongs to the meta-optimization 
methodology and thus can be applied in a broad research area. Original optimization problem can simply be 
extended to allow spatially and temporally varying environmental condition so that each local
entity has a different but closely related fitness function. 
Genetic entities can then mate with each other, locally and remotely, which constitute the 
properly arranged heterogeneity in genetic system. 
For example, one can combine the environmental condition of our model with 
the crossover scheme of meiosis GA\cite{meiosis} or gendered GA\cite{gendered}, or the migration process of island model\cite{island}, with 
small modification of the objective function.

We argue that such wide applicability of our methodology ensures its potential practicality. 
Although the current framework of our GA is not appropriate to compare the performance with 
existing algorithms for a 3D spin-glass system, we expect that the central concept of our method is easily applicable to 
other existing methods to further enhance the performance.
We conclude that different environmental conditions can be practically helpful
in finding a better optimum in a shorter time, by allowing a
certain level of genetic diversity in a gene pool. We plan to investigate
other optimization problems by applying our suggested algorithm in the near future.

\section*{Acknowledgments}
This work was supported by the National Research Foundation of Korea (NRF) 
grant funded by the Korea government (MSIT) Grant No. 2019R1A2C2089463.

%\bibliographystyle{elsarticle-num}
%\bibliography{refer.bib}

\end{document}